\documentstyle[12pt,aaspp4]{article}
\input epsf
\begin{document}

\title{Astrometric Microlensing of Distant Sources due to the Stars in the Galaxy}
\author{Mareki Honma\altaffilmark{1,2} \& Tomoharu Kurayama\altaffilmark{1,3}}
\altaffiltext{1}{VERA Project Office, National Astronomical Observatory of Japan, Mitaka 181-8588, Japan}
\altaffiltext{2}{Earth Rotation Division, National Astronomical Observatory of Japan, Mizusawa 023-0861, Japan}
\altaffiltext{3}{Department of Astronomy, University of Tokyo, 7-3-1 Hongo, Bunkyo-ku, Tokyo 113-0033, Japan}
\authoremail{honmamr@cc.nao.ac.jp}
\slugcomment{submitted to the Astrophysical Journal}

\begin{abstract}
We investigate properties of astrometric microlensing of distant sources (such as QSOs and radio galaxies) caused by stars in the Galaxy, mainly focusing on application to the VERA (VLBI Exploration of Radio Astrometry) project.
Assuming typical parameters for the Galaxy disk and bulge, we show that the maximum optical depth for astrometric shift of 10 $\mu$as-level is $8.9\times 10^{-2}$ for QSO-disk lensing case and $3.8\times 10^{-2}$ for QSO-bulge lensing case.
We also find that the maximum optical depth for QSO-disk lensing is larger by an order of magnitude than that for disk-disk or bulge-disk lensing case (assuming a typical source distance of 8 to 10 kpc).
In addition to optical depth, we also calculate the event rate and find that the maximum event rate for QSO-disk lensing case is 1.2$\times 10^{-2}$ event per year, which is about 30 times greater than that for disk-disk lensing case.
This high event rate implies that if one monitors 10 QSOs behind the Galactic center region for 10 years, at least one astrometric microlensing event should be detected.
Therefore, monitoring distant radio sources with VERA can be a new tool to study astrometric microlensing caused by stars in the Galaxy.

We also study the event duration of astrometric microlensing, and find that the mean event duration for QSO-disk lensing is 7.5 yr for QSOs located near the galactic center.
This event duration for QSO-disk lensing is reasonably short compared to the project lifetime of VERA, which is anticipated to be $\sim$ 20 yr.
We also find that while the minimum event duration for bulge-bulge lensing is as short as 2.6 yr, the event duration for disk-disk lensing case cannot be shorter than 15 yr. 
Thus, although astrometric microlensing of bulge sources/lenses can be studied by optical astrometric missions like SIMA and GAIA, detections of disk events with the space astrometric missions are fairly difficult because of the limited project lifetime (typically $\sim$5 yr) as well as the heavy dust extinction.
Therefore, for studying astrometric microlensing by disk stars, VERA can be one of powerful tools by observing distant sources like QSOs and radio galaxies.
We discuss the implications of astrometric microlensing for VERA by focusing on estimating the lens mass, and also present some possible candidates of radio sources toward which astrometric microlensing events should be searched for with VERA.

\end{abstract}
\keywords{Galaxy : disk --- Galaxy : structure --- gravitational lensing --- stars : low-mass, brown dwarfs}

\section{Introduction}

Gravitational microlensing is one of the most promising tools to study invisible lenses like MACHOs and low-mass stars in the Galaxy.
Paczynski (1986) first proposed to use microlensing effect to search for MACHOs in the Galaxy's halo based on a photometric monitoring of millions of stars in the Galaxy's bulge and the Magellanic Clouds.
A number of groups have been conducting such `photometric' searches of microlensing to detect source magnification by gravitational microlensing (e.g., MACHO, EROS, OGLE, MOA, etc.), finding more than a hundred of photometric microlensing events.

In addition to such photometric microlensing, there is another type of microlensing called `astrometric microlensing', in which the positional shift of the lensed image, rather than the magnification of the source, is used to detect microlensing events.
Recently, a number of studies have been made on such `astrometric' microlensing events (e.g., Hosokawa et al. 1993; 1997; H\o{}g, Novikov \& Polnarev 1995; Miyamoto \& Yoshii 1995; Walker 1995; Paczynski 1996; 1998; Miralda-Escude 1996; Boden et al. 1998; Dominik \& Sahu 2000).
One of the major findings of these studies is that the probability of astrometric microlensing is much larger than that of photometric microlensing (e.g., Miralda-Escude 1996; Hosokawa et al.1997; Dominik \& Sahu 2000; Honma 2001).
For instance, the size of a photometric lens is equal to the Einstein-ring size given by
\begin{equation}
\label{eq:E-ring}
R_{\rm E}=\sqrt{\frac{4GM}{c^2} \frac{D_{\rm d}D_{\rm ds}}{D_{\rm s}}}.
\end{equation}
Here $M$ is the lens mass, and $D_{\rm d}$, $D_{\rm ds}$, and $D_{\rm s}$ are the observer-lens distance, lens-source distance, and observer-source distance, respectively.
If a source comes within this radius from the lens, a source is magnified by more than a factor of 1.34.
Meanwhile, the size of an astrometric lens is larger than $R_{\rm E}$ by a factor of $\beta_{\rm max}$, which is given by following equation (e.g., Honma 2001).
\begin{equation}
\label{eq:beta_max}
\beta_{\rm max} = \theta_{\rm E}/\theta_{\rm min}.
\end{equation}
Here $\theta_{\rm E}$ is the angular size of the Einstein-ring radius ($\theta_{\rm E}\equiv R_{\rm E}/D_{\rm d}$), and $\theta_{\rm min}$ is the minimum angular shift that can be detected by astrometric observation.
Thus, if a source comes within $\beta_{\rm max} R_{\rm E}$ from the lens, the source position is shifted by larger than $\theta_{\rm min}$.
For astrometric lensing, $\beta_{\rm max}$ can be as large as 50 $\sim$ 100 assuming $\theta_{\rm min}$ of 10-$\mu$as level (e.g., Honma 2001).

Such a high astrometric accuracy, although not yet achieved, will be available soon because a number of astrometric missions are planned in early 21st centuries.
For instance, there are four space astrometric missions to be launched by $\sim$2010: DIVA, FAME, SIM and GAIA.
Astrometric accuracies anticipated for those missions are: 200 $\mu$as for DIVA, 50 $\mu$as for FAME, and 10 $\mu$as (or possibly higher) for SIM and GAIA.
In addition to those space missions, there is another ground-based astrometric mission called VERA (VLBI Exploration of Radio Astrometry, e.g., Sasao 1996; Honma, Kawaguchi \& Sasao 2000 and reference therein), which utilize a phase-referencing VLBI technique for astrometry of radio sources.
The most remarkable difference between VERA and space astrometric missions is that VERA can observe thousands of distant sources like QSOs and radio galaxies to trace effect of astrometric microlensing.
An advantages of using distant radio sources is that the column density of lens can be much higher, leading to higher event probability.

Investigating disk stars by astrometric microlensing with missions described above has significant scientific merits.
First, astrometric microlensing provide us a new tool to study the lower end of stellar mass function, which is thought to be dominant populations in the Galaxy's disk.
To date, the shape of the mass function remains uncertain in particular at its lower end, because the lower main-sequence stars are too faint for optical observations.
Studying the low-mass star may also inform us about the nature of dark matter in the disk, as the disk dark matter may be baryonic matter in the form of low-mass stars and/or brown dwarfs.
Further, studies of astrometric microlensing by disk stars will also provide fundamental information on parameters of the Galaxy's disk.
For instance, one can estimate the disk scale length from the optical depth distribution with the Galactic latitude $l$ (e.g., Dominik \& Sahu 2000).
Also, it is possible to extract the disk density from astrometric microlensing events provided sufficient number of events are observed for statistics.
From the scale length and disk density, one can estimate the total mass of the Galaxy's disk, and thus the density profile of dark halo inside in the disk region can be also deduced.

While the astrometric microlensing of stars are extensively studied in previous studies focusing on implications for SIM and GAIA(e.g., Dominik \& Sahu 2000), there has been no study on astrometric microlensing of distant radio sources by disk stars considering an application to VERA.
The most remarkable aspect of using distant sources is that some important parameters such as the lens mass may be determined independently of lens distance (e.g., Honma 2001), as well as high event probability due to large column density toward a source.
For these reasons, in the present paper we extensively study the astrometric microlensing of distant sources by disk stars, and discuss its implications for VERA.

The plan of this paper is as follows: in section 2 we present calculations of optical depth, and in section 3 event rate.
We also compare results for QSO-disk/bulge lensing cases with disk/bulge-disk lensing cases.
In section 4, we present event durations of astrometric events, and in section 5 we discuss how the lens mass can be estimated from event durations.
Finally in section 6, we discuss the implication of astrometric microlensing for VERA, and also summarize the findings of the present paper.

\section{Optical Depth}

\subsection{basic equations}

The optical depth of astrometric microlensing event is defined as follows (e.g., Miralda-Escude 1996):
\begin{equation}
\label{eq:tau_ast}
\tau_{\rm ast} = \int \pi \beta_{\rm max}^2 R_{\rm E}^2\; \frac{\rho}{M}\; dD_{\rm d}.
\end{equation}
Here, $R_{\rm E}$ is the Einstein-ring radius, $\rho$ is the mass density in the lens, $D_{\rm d}$ is the lens distance from the observer, and $M$ is the lens mass.
Note that this lens mass should be regarded as the mean lens mass averaged over the possible range of stellar mass function (the effect of stellar mass function will be discussed later in the present paper).
The factor $\beta_{\rm max}$, defined in equation (\ref{eq:beta_max}), denotes the ratio of astrometric lens size to the Einstein ring size.
The factor of $\beta_{\rm max}^2$ in equation (\ref{eq:tau_ast}) is crucial to the difference of the optical depth between astrometric and photometric microlensing.
As is mentioned earlier, in the distant source case $\beta_{\rm max}$ could be as large as 50 $\sim$ 100, and thus the optical depth for astrometric microlensing could be larger than that of photometric microlensing by 3 to 4 orders of magnitudes.
By substituting the expression of $R_{\rm E}$ and $\beta_{\rm max}$ into equation (\ref{eq:tau_ast}), one can obtain following expression of the optical depth,
\begin{equation}
\label{eq:tau_ast2}
\tau_{\rm ast} = \frac{16\pi G^2 M}{c^4 \theta_{\rm min}^2} \int \rho\; \left(1-\frac{D_{\rm d}}{D_{\rm s}}\right)^2\; dD_{\rm d}.
\end{equation}
Interestingly, the optical depth is proportional to the mean lens mass $M$, in contrast to photometric microlensing in which the optical depth is independent of the mean lens mass $M$.

\subsection{mass model}

For the density distribution in the Galaxy's disk, we assume an exponential disk for both radial and vertical profiles, namely,
\begin{equation}
\label{eq:rho-disk}
\rho_{\rm d}(R,z) = \rho_0 \exp \left(-\frac{R-R_0}{d}-\frac{|z|}{h}\right).
\end{equation}
Here, $\rho_0$ is the disk density in the vicinity of the Sun, and $d$ and $h$ are the radial and vertical scale lengths, respectively.
We take rather conservative values for these parameters as $\rho_0=0.08 M_\odot$ pc$^{-3}$, $d=3.5$ kpc, and $h=300$ pc (table 1 summarizes the model parameters used in this paper).
Note that the disk model and its parameters considered here are the same as those used in previous studies (e.g., Hosokawa et al. 1997; Dominik, Sahu 2000; Honma 2001).

In order to take into account the contributions from bulge stars, we consider a bulge model whose density distribution is given by a spherical Plummer model.
The density distribution of the bulge is given by
\begin{equation}
\rho_{\rm b}(r) = \left(\frac{3M_{\rm b}}{4\pi a^3}\right)\left(1+\frac{r^2}{a^2}\right)^{-\frac{5}{2}},
\end{equation}
where $M_{\rm b}$ is the total mass and $a$ is the scale length of the bulge, respectively.
Although the bulge may be elongated and bar-like system as suggested by recent studies, here we do not consider such an asymmetric density distribution, as the precise parameters of bulge elongation is still highly uncertain.
For the bulge parameters, we assume the bulge mass $M_{\rm b}$ of $0.8\times 10^{10} M_\odot$ and the scale length $a$ of 1 kpc (e.g., Dehnen \& Binney 1998).

Here we also introduce the stellar mass function to calculate the mean lens mass $M$.
To be conservative, we assume the Salpeter-type stellar mass function with the power law index of 2.35, namely,
\begin{equation}
\phi (m) = C_\phi m^{-2.35} \;\;\;({\rm for}\;\;\; m_{\rm L}\le m \le m_{\rm U}).
\end{equation}
Here $C_\phi$ gives a scaling constant, and $m_{\rm L}$ and $m_{\rm U}$ are the lower and upper cutoff of the stellar mass function.
Based on that mass function, we obtain the mean lens mass as
\begin{equation}
M = \frac{\int\; m \phi (m)\; dm}{\int\; \phi (m)\; dm}.
\end{equation}
With a lower cutoff $M_{\rm L}$ of 0.08$M_\odot$, which corresponds to the hydrogen burning limit, we obtain the mean mass of $0.3M_\odot$ (note that $M$ is independent of $M_{\rm U}$ provided $M_{\rm U}$ is sufficiently large).
This mass $M$ of $0.3M_\odot$ corresponds to that of lower main-sequence stars, which are thought to be dominant populations in the Galaxy's stellar component.
Thus, in the following sections we assume the mean lens mass of 0.3$M_\odot$ except otherwise noted.
Of course, the mean mass is quite uncertain since we have little knowledge of the lower end of the stellar mass function.
In case of different mean mass, the optical depth for mean lens mass $M'$ can be obtained by multiplying a factor of $(M'/M)$ to the optical depth obtained in the present paper.

\subsection{results}

Using equations described above, one can obtain the optical depth for astrometric microlensing.
In the present paper, we consider several cases with different source and lens populations.
Table 2 summarizes the cases we consider in the present paper.
For QSO-disk and QSO-bulge lensing, in which distant QSO is being lensed by disk or bulge stars, the distance to QSO is assumed to be infinite.
On the other hand, for disk-disk lensing case, we assume the source distance $D_{\rm s}$ of 8 kpc.
The first reason for setting $D_{\rm s}=8$ kpc is that the distance corresponds to that to the disk stars near the Galactic center, where the disk density is highest.
The second reason is that bright stars like AGB have $M_V$ of 0 mag, and hence their apparent magnitude becomes about 15 mag at the distance of 8 kpc in case of no interstellar extinction.
Since the stars at low galactic latitude significantly suffer from strong absorption by the interstellar medium (later we will show that we have to observe stars at fairly low latitude to detect an astrometric microlensing), their apparent magnitude could be reduced to $\sim$20 mag or even fainter, which is close to or below the limiting magnitude of optical astrometric missions like SIM and GAIA.
Meanwhile, for bulge-disk or bulge-bulge lensing case, we assume a slightly longer distance, $D_{\rm s}$ of 10 kpc.
This is because a part of bulge sources can be seen through low-extinction regions like the Baade window, which are a few degrees away from the Galactic center.
In these windows, we may be able to observe slightly far sources compared to disk sources partly because the extinction is low, and partly because there are quite many bright stars like bulge red giants (note that these {`}windows{'} are not suitable for search for disk lens event since their galactic latitude is too high for disk lensing; for instance, $b\sim 4^\circ$ for the Baade window).

Figure 1 shows the optical depths calculated for the cases described above.
First of all, for QSO-disk/bulge lensing case, the maximum optical depth is obtained for sources behind the Galactic center (i.e., $(l,\, b)=(0^\circ,\, 0^\circ$)), being $8.9\times 10^{-2}$ for disk lens, and $3.8\times 10^{-2}$ for bulge lens, respectively.
This result indicates that nearly one of ten sources in this direction is always lensed by disk stars, and the optical depth of bulge lens is nearly half of that of disk lens (we will discuss how to discriminate disk lens event from bulge lens event later).
However, the optical depth for bulge lens rapidly decrease with increasing $l$, and its contribution becomes negligible at $l\ge 10^\circ$.
For QSO-disk lensing, the dependence on $b$ is much stronger that on that $l$.
For instance, the optical depth at $(l,\, b)=(0^\circ,\, 5^\circ)$ is smaller than that at $(l,\, b)=(0^\circ,\, 0^\circ)$ by almost an order of magnitude.
This strong dependence comes from the short vertical scale length $h$ (300 pc) compared to the radial scale length $d$ (3.5 kpc).
These results indicate that one should look for astrometric microlensing by disk stars at fairly low Galactic latitude (e.g., less than 2 degrees).

For comparison, we also show in figure 1 the optical depth for disk-disk lensing case and bulge-bulge/disk lensing cases.
From figure 1 one can find that the maximum optical depth for disk-disk lensing is 6.2$\times 10^{-3}$, being smaller by a factor of 15 than the maximum optical depth for QSO-disk lensing.
Also, the maximum optical depth for bulge-disk lensing case is 9.3$\times 10^{-3}$, 10 times smaller than that for QSO-disk lensing.
The optical depth for bulge-bulge lensing case is far smaller, being 1.0$\times 10^{-3}$.
These significant differences of optical depths between distant source and disk/bulge source cases can be understood if one carefully looks at the equation (\ref{eq:tau_ast2}).
In the distant source case, the factor $(1-D_{\rm d}/D_{\rm s})^2$ is almost unity within the entire Galaxy, and thus the largest contribution to the optical depth occurs at the galactic center, where the disk density is highest.
On the other hand, in the disk-disk lensing case the largest contribution to the optical depth comes from stars near the Sun (where the disk density is relatively low), because the factor of $(1-D_{\rm d}/D_{\rm s})^2$ becomes fairly small for lenses near the Galactic center.
This fact also explains that for the disk-disk lensing case the dependence of $\tau_{\rm ast}$ on $l$ is not so strong as that for the QSO-disk case: the density of local stars, which contribute most to the optical depth, does not vary strongly with changing $l$.

\section{Event Rate}

\subsection{basic equation}

The event rate is another important quantity in microlensing study, since the event rate is directly related to the observational strategy.
The event rate is usually defined as the number of lenses coming inside the microlensing tube per unit time (e.g., Griest 1991).
The radius of the microlensing tube is given by the lens size, which is $\beta_{\rm max} R_{\rm E}$ for astrometric microlensing.
By definition, the event rate strongly depends on the distribution of tangential velocity of lenses, and thus is usually a function of velocity distribution of lens population.
In this section, for simplicity, we assume that the tangential velocity of lens relative to the tube $v_\perp$ is uniquely determined at a given distance $D_{\rm d}$.
This $v_\perp$ corresponds to an averaged tangential velocity at a given distance, which can be obtained by integrating the velocity distribution function over the possible velocity range (in later section we will treat the effect of velocity distribution function more precisely).
Under this assumption, the event rate for astrometric microlensing $\Gamma_{\rm ast}$ can be written as
\begin{equation}
\label{eq:gamma}
\Gamma_{\rm ast} = \int 2 \beta_{\rm max} R_{\rm E}\; \frac{\rho}{M} v_\perp\; dD_{\rm d}.
\end{equation}
This expression is correct regardless of the shape of velocity distribution function.
For instance, in case that the lens population has an isotropic velocity distribution, the number flux of lenses which crosses unit area per unit time ($F$) is given by
\begin{equation}
F_{\rm iso} = \frac{1}{\pi}\; \frac{\rho}{M}\; v_\perp.
\end{equation}
Note that only the population coming in the tube is considered but the population going out of the tube is ignored.
Since the area of the lensing tube per unit length $dD_{\rm d}$ is $2\pi \beta_{\rm max} R_{\rm E} dD_{\rm d}$, one can obtain the event rate per unit length ($d\Gamma_{\rm ast}/dD_{\rm d}$) as
\begin{equation}
\label{eq:dgamma}
\frac{d\Gamma_{\rm ast}}{dD_{\rm d}} = 2\pi \beta_{\rm max} R_{\rm E} \times \frac{1}{\pi}\; \frac{\rho}{M}\; v_\perp = 2 \beta_{\rm max} R_{\rm E}\; \frac{\rho}{M}\; v_\perp,
\end{equation}
which is a differential form of equation (\ref{eq:gamma}).
On the other hand, if the lens population is perfectly streaming in a direction without any random motion, the number flux of lenses per unit area per time is given by $\rho/M v_\perp$ (assuming that the streaming motion is perpendicular to the unit area considered).
Since the geometric cross section of the tube per unit length seen from the lens population is $2 \beta_{\rm max} R_{\rm E} dD_{\rm d}$, one may obtain the same expression for the differential event rate $d\Gamma_{\rm ast}/dD_{\rm d}$ for the isotropic case.

Substituting equations (\ref{eq:E-ring}) and (\ref{eq:beta_max}) into (\ref{eq:gamma}), one may obtain the following expression of the event rate,
\begin{equation}
\Gamma_{\rm ast} = \frac{8 G}{c^2 \theta_{\rm min}} \int \rho\; v_\perp\; \left(\frac{D_{\rm ds}}{D_{\rm s}}\right) dD_{\rm d}.
\end{equation}
Note that the event rate for astrometric microlensing is independent of the lens mass $M$, as is pointed out in previous studies (e.g., Dominik \& Sahu 2000; Honma 2001).

\subsection{velocity model}

The tangential velocity of the lens relative to the tube can be defined using the motion of source, lens and observer as follows,
\begin{equation}
\vec{v}_\perp = \frac{D_{\rm ds}}{D_{\rm s}}\vec{v}_{{\rm o} \perp} + \frac{D_{\rm d}}{D_{\rm s}} \vec{v}_{{\rm s} \perp} - \vec{v}_{{\rm d} \perp}.
\end{equation}
Here $\vec{v}_{{\rm o} \perp}$, $\vec{v}_{{\rm s} \perp}$, and $\vec{v}_{{\rm d} \perp}$ are the two dimensional velocities of the observer, source, and lens perpendicular to the line of sight.
In order to take into account the effect of velocity dispersion in a rather simple manner, here we calculate the lens tangential velocity $v_\perp$ as
\begin{equation}
\label{eq:v_tan}
v_\perp = \sqrt{|\vec{v}_\perp|^2 + \sigma_{\rm lens}^2 + \left(\frac{D_{\rm d}}{D_{\rm s}}\right) \sigma_{\rm source}^2},
\end{equation}
where $\sigma_{\rm lens}$ and $\sigma_{\rm source}$ are the velocity dispersion of the lens and source population, respectively.

The velocities of disk stars can be easily described based on the so-called flat-rotation model.
Since the rotation velocity of the Galaxy's disk is known to be almost constant at any galacto-centric distance, here we assume a constant circular velocity $v_{\rm circ}$ of 220 km/s (e.g., Kerr \& Lynden-Bell 1986).
Given the circular velocity in the disk and star position ($l$, $b$ and the distance to the star), one can determine its tangential motion with respect to the line of sight caused by the galactic rotation.
In addition to circular rotation velocity, stars in the disk are known to have velocity dispersion of 15 $\sim$ 20 km/s, and thus in case of disk stars (acting as both lens and source) we set $\sigma_{\rm lens}$ and/or $\sigma_{\rm source}$ of 20 km/s.

In figure 2 we show the distribution of $v_\perp$ with the lens distance $D_{\rm d}$.
Figure 2 clearly demonstrates the difference of tangential velocities in two cases; QSO-disk lensing and disk-disk lensing case.
For the QSO-disk lensing case, the largest tangential velocity is obtained toward a source behind the Galaxy's center, which is as large as 440 km/s.
This velocity, corresponding to 2$\times v_{\rm circ}$, is obtained because behind the Galactic center the lens motion becomes completely opposite to the observer motion.
On the other hand, for the disk source case the largest tangential velocity is not larger than 100 km/s.
This is due to the fact that the source is not further than the Galaxy's center, and thus the source as well as the lens are moving in the same direction as that of the observer.
In particular, in the vicinity of the Sun (i.e., within 1 kpc from the Sun), where the contribution to the optical depth becomes the maximum for disk-disk lensing case, the tangential velocity $v_\perp$ is no larger than 40 km/s, which is an order of magnitude smaller than the largest tangential velocity for QSO-disk lensing case.
These facts imply that the event rate for disk source case could be significantly smaller than that for distant source case.

On the other hand, for the bulge model we use the isothermal velocity dispersion obtained from the Plummer potential.
For the tangential velocity dispersion, the Plummer model gives (e.g., Lynden-Bell 1962)
\begin{equation}
\sigma^2 = \frac{1}{3}\frac{GM_{\rm b}}{\sqrt{r^2+a^2}}.
\end{equation}
The parameters $M_{\rm b}$ and $a$ are the bulge mass and the scale length, and taken to be $M_{\rm b}=0.8\times 10^{10}M_\odot$ and $a=1$ kpc, as described in the previous section.
With these parameters, the tangential velocity dispersion at the Galactic center becomes $\sim 110$km/s, which is in agreement with observed velocity dispersion (e.g., Tiede \& Terndrup 1999).
The bulge is also known to have a rigid rotation with an angular velocity of $\sim 50$ km/s/kpc (Tiede \& Terndrup 1999).
In the present paper, however, we do not include this rigid rotation, since at the central bulge (where most of bulge lensing events occur), the tangential velocity is dominated by the velocity dispersion rather than the rotational motion.

\subsection{results}

Based on the tangential velocity and density distribution described above, one can calculate the event rate of astrometric microlensing using equation (\ref{eq:gamma}).
Figures 3 shows the event rate distributions for cases considered in the previous sections (see table 2).
From figure 3 one sees that for QSO-disk lensing case the maximum event rate is $1.2\times 10^{-2}$ event/yr, which is reasonably high for practical observations.
In fact, this indicates that an event can be detected if one observes 100 sources for a year, or 10 sources for 10 years.
If 50 sources are monitored for the whole mission lifetime of VERA ($\sim 20$yr), one can expect to find $\sim$10 astrometric microlensing events.
For QSO-bulge lensing case, the maximum event rate is $5.0\times 10^{-3}$, nearly half of that for QSO-disk lensing.
Thus, for QSO behind the galactic center, the total event rate will be around 1.7$\times 10^{-2}$ event per year.

The fact that QSO-disk lensing and QSO-bulge lensing rates are comparable raises a new problem, that is how to discriminate disk lens events from bulge lens events.
To do this, one can use an additional observable of astrometric microlensing event, which is the direction of lens motion.
If the whole source motion along the circular trajectory is monitored, then one can directly determine the direction of lens motion; the lens motions is perpendicular to the line connecting between the original source position and the position of maximum image shift.
For QSO-disk lensing, one can expect that the direction of the lens motion is along the galactic plane.
For instance, a typical disk lens behind the galactic center is moving at 400 km/s with respect to the lensing tube, while that star is likely to have a random motion of only 20 km/s or so (disk velocity dispersion).
Therefore, the alignment of the lens direction and the galactic plane should be within a few degrees ($\sim$ 20/400 radian).
On the other hand, since for QSO-bulge lensing case the lens tangential velocity is dominated by the random motion of bulge stars, there should be no tight correlations between the direction of the lens motion and the galactic plane.

For disk-disk lensing case the largest event rate is $4.0\times 10^{-4}$, being $\sim 30$ times smaller than that for the QSO-disk lensing case, and also 10 times smaller than that for QSO-bulge case.
At the galactic center ($l$, $b$)=($0^\circ,\, 0^\circ$) the event rate for disk-disk lensing is only $1.3\times 10^{-4}$.
This small event rate comes from the small tangential velocity for the disk source case as well as the fact that stars around the Galactic center, where the stellar density is highest, make little contributions to the disk-disk lensing (as discussed for the optical depth calculation).
At the galactic center the bulge-disk lensing case and bulge-bulge case give higher event rate than that for the disk-disk lensing case, being $1.3\times 10^{-3}$ and $4.0\times 10^{-4}$, respectively.
This is due to higher velocity dispersion in the bulge than in the disk.
Hence, if one looks for astrometric microlensing events with space astrometric missions like SIM and GAIA, it is better to observe bulge sources rather than disk stars.

Note that for disk-disk lensing cases the maximum event rate is obtained at $l\sim 10^\circ$ rather than toward the Galactic center.
This is because the event rate (equation[\ref{eq:gamma}]) is obtained by integrating a product of the density $\rho$ and tangential velocity $v_\perp$:
while the maximum density along the line of sight is decreasing with $l$, the maximum tangential velocity is peaked around $l\sim 10^\circ$.
On the other hand, toward $l=0^\circ$ the rotation velocities of the source, lens and observer cancel out, and the tangential velocity reduces to the velocity dispersion $\sigma$, making the event rate fairly small.

\section{Event Duration}

\subsection{basic equations}

The event duration, which is one of the most important observables in individual microlensing events, is defined as period during which a lens is located within a microlensing tube.
For astrometric microlensing, the event duration can be described as
\begin{equation}
\label{eq:t-ast}
t_{\rm ast} = \frac{2\beta_{\rm max} R_{\rm E}}{v_\perp} \cos \delta,
\end{equation}
where $2\beta_{\rm max} R_{\rm E}$ corresponds to the lens diameter and the angle $\delta$ defines the lens-source impact parameter as $\beta_{\rm max} R_{\rm E}\times \sin \delta$.
The averaged event duration can be obtained by integrating $t_{\rm ast}$ over possible range of $\delta$ as (e.g., Honma 2001)
\begin{equation}
\label{eq:t-ast-av}
\bar{t}_{\rm ast} = \frac{\pi}{2}\left(\frac{\beta_{\rm max} R_{\rm E}}{v_\perp}\right).
\end{equation}
Substituting equations (\ref{eq:E-ring}) and (\ref{eq:beta_max}) into (\ref{eq:t-ast-av}), one can obtain 
\begin{equation}
\label{eq:t-ast-av2}
\bar{t}_{\rm ast} = \frac{2\pi GM}{c^2 v_\perp \theta_{\rm min}}\left(\frac{D_{\rm ds}}{D_{\rm s}}\right).
\end{equation}
Note that the average event duration is defined at any lens distance $D_{\rm d}$, and varies with changing $D_{\rm d}$ through the factor of $(D_{\rm ds}/D_{\rm s})$ as well as $v_\perp$, which also depends on $D_{\rm d}$.
Thus, in order to obtain expected event duration toward a source (hereafter denoted as $\langle t_{\rm ast}\rangle$), one has to integrate the average event duration $\bar{t}_{\rm ast}$ along the line of sight from observer to source.
We also note that for distant source case, the averaged event duration becomes independent of the lens distance as $(D_{\rm ds}/D_{\rm s})$ approaches to unity (Honma 2001).
Hence, for distant source case, the uncertainty in lens mass determination can be smaller than that for disk source case.

While one can obtain the expected event duration $\langle t_{\rm ast}\rangle$ by integrating equation (\ref{eq:t-ast-av2}) over possible range of $D_{\rm d}$, here we use another simple way, that is to use the optical depth and event rate obtained in the previous sections.
These two quantities are related to the expected event duration through the following equation,
\begin{equation}
\label{eq:t-ast-expected}
\langle t_{\rm ast} \rangle = \frac{\tau_{\rm ast}}{\Gamma_{\rm ast}}.
\end{equation}
Based on this equation together with the optical depth and event rate calculated in previous sections, we calculate the event duration.

\subsection{results}

Figure 4 shows the distribution of $\langle t_{\rm ast}\rangle$ with various lines of sight, for both of disk/bulge source and distant source cases.
For QSO-disk lensing, $\langle t_{\rm ast}\rangle$ is found to be 7.5 yr toward the Galactic center (in case of $b=0^\circ$).
Around the Galactic center region (e.g., $l\le 30^\circ$), the expected event duration changes only slightly with $l$, but increases rapidly for the outer Galaxy lenses ($l=90^\circ$ to 180$^\circ$), reaching at 91 yr at the anti-galactic center.
Hence, in order to detect events within rather practical time-scale (i.e., less than 20 yr), one should observe sources around the Galactic center regions.

 For QSO-disk lensing toward the galactic center region, the lenses may be divided into two populations based on their location along the line of sight, namely the lenses in front of the galactic center and the lenses behind the galactic center.
For these two populations, the relative proper motions of the lens are completely different: while the proper motion of lens in front of the galactic center is dominated by the velocity dispersion in the disk, the proper motion of lens behind the galactic center is dominated by the galactic rotation (see figure 2).
Due to this difference, these two populations also have different distribution of event rate and event durations.
To show this clearly, in figure 5 and 6 we show the dependence of differential event rate and event duration on the lens distance.
As seen in figure 5, most of events are likely to be caused by the stars behind the galactic center because of the large proper motion.
For instance, toward $l=0^\circ$ about 95\% of events are caused by the stars behind the galactic center.
Figure 6 shows that the events caused by stars behind the galactic center have shorter event duration because of larger proper motion.
In fact, for the case of $l=0^\circ$, the lenses behind the galactic center gives a typical duration of 4.1 yr, while the lenses in front of the galactic center gives a duration of 91 yr.
In practice, however, the distinction between two populations is not as clear as in figure 5 or 6 because of the effect of the  mass function or velocity dispersion.
For instance, a less massive star in front of the galactic center could cause shorter event, which may be comparable to events caused by more massive star behind the galactic center (in the next section, we will discuss the effect of the mass function and the velocity dispersion more precisely).

On the other hand, for disk-disk lensing case, the dependence of $\langle t_{\rm ast}\rangle$ on $l$ (figure 4) is not simple, having multiple local maximum and minimum values.
The minimum event duration is obtained toward $l\sim 10^\circ$ where the tangential velocity of the lens is largest (see figure 2).
The minimum event duration is found to be 15 yr, being twice of the minimum duration for the distant source case.
With increasing the Galactic longitude $l$, the expected event duration increases toward $l\sim 90^\circ$, where the tangential velocity is dominated by the disk velocity dispersion.
Toward the outer Galaxy ($l=90^\circ$ to $180^\circ$), the duration is between 55 to 70 yr depending on $l$, with another local maximum at $l=180^\circ$.
Note that for the disk star case the tangential velocity becomes equal to the disk velocity dispersion at both $l=0^\circ$ and $180^\circ$, but the expected event duration $\langle t_{\rm ast}\rangle$ is slightly different from each other, since the mean distance to the lens, and hence the mean angular size of the lens are different at $l=0^\circ$ and $180^\circ$ (this difference of mean lens distance comes from the density distribution in the disk).
In any case, the event duration cannot be less than 10 yr for the disk source case.
Hence the event duration seems too long for practical observation when one considers that the lifetime of space astrometric missions is $3\sim 5$ yr at longest.

Note that this long duration for disk-disk event mainly comes from the assumption that the source is no further than the galactic center (the distance is assumed to be 8 kpc in this paper).
In such a disk-disk event, observer, lens and sources move toward nearly same direction, and thus the lens-source proper motion is small, being dominated by the disk velocity dispersion.
On the other hand, for a source star behind the galactic center, the relative proper motion of lens is dominated by the galactic rotation rather than the disk velocity dispersion (see figure 2).
Thus, if one can observe a distant disk star located behind the galactic center, one would be able to detect disk-disk events with relatively short duration.
However, in practice, this is extremely difficult for astrometric space missions like SIM and GAIA because of heavy dust extinction in the galactic plane.
To detect such an event, an astrometric mission at infrared will be necessary.

For the bulge-source cases (bulge-bulge and bulge-disk lensing), figure 4 shows that the event duration can be practically short.
For instance, at the galactic center the expected event duration is 8.1 yr for bulge-disk lensing case, and 2.6 yr for bulge-bulge lensing.
This short event duration comes from large velocity dispersion of bulge stars.
In case of bulge-bulge lensing, there is another reason for short event duration, that the lens size itself is substantially small because of short lens-source distance (since both lens and source are in the bulge).
Therefore, if one looks for astrometric microlensing at optical/near infrared bands, which are the main observational bands of space astrometric missions, it is practical to monitor the bulge sources, rather than disk sources.

\section{Effect of Mass Function and Velocity Dispersion}

When an astrometric microlensing event is detected, the lens mass can be estimated from the event duration, because the event duration is proportional to the lens mass in astrometric microlensing (see, for instance, equation [\ref{eq:t-ast-av2}]).
As is mentioned earlier, the most interesting aspect of equation (\ref{eq:t-ast-av2}) is that for distant source case the factor $D_{\rm ds}/D_{\rm s}$ approximately becomes unity, and hence the average event duration becomes independent of the lens distance.
We note, however, that one cannot uniquely determine the lens mass for individual events, since there is no way to know the lens-source impact parameters and/or tangential velocity for individual events.
Also, disk/bulge stellar masses are not unique but have a certain range according to the stellar mass function.
Therefore, we need some statistical treatments when converting the event duration to the mean lens mass.
In this section, we consider such effects based on more realistic models of the velocity dispersion and stellar mass function in the bulge and disk.

Instead of equation (\ref{eq:gamma}), here we consider more general form of the event rate given by
\begin{equation}
\Gamma_{\rm ast} = \int\int\int\int \beta_{\rm max} R_{\rm E}\; \frac{\rho}{m} v_\perp\; f(v_\perp)\; \phi(m) \; \cos \delta \; d\delta\, dv_\perp\, dm\, dD_{\rm d}.
\end{equation}
Here $f(v_\perp)$ expresses the distribution of lens tangential velocity relative to the microlensing tube, and $\phi(m)$ denotes the stellar mass function, which are normalized so that $\int f(v_\perp)\; dv_\perp=1$ and $\int \phi(m)\; dm=1$, respectively.
From the relation between $v_\perp$ and $t_{\rm ast}$ (equation \ref{eq:t-ast}), one can obtain 
\begin{equation}
dv_\perp = - \frac{2 \beta_{\rm max} R_{\rm E}}{t_{\rm ast}^2} \cos\delta \, dt_{\rm ast}.
\end{equation}
Together with above equations, the event rate may be written as
\begin{equation}
\label{eq:gamma-with-fm}
\Gamma_{\rm ast} = \frac{256 G^3}{c^6 \theta_{\rm min}^3} \int\int\int\int \frac{\rho m^2}{t_{\rm ast}^3} \left(\frac{D_{\rm ds}}{D_{\rm s}}\right)^3\; f\; \phi\; \cos^3\delta\; d\delta\, dt_{\rm ast}\, dm\, dD_{\rm d}
\end{equation}
By differentiating equation (\ref{eq:gamma-with-fm}) with respect to $t_{\rm ast}$, one can also obtain the differential event rate $d\Gamma_{\rm ast}/dt_{\rm ast}$, which shows the distribution of observed event rate against event duration.

Figure 7 shows this $d\Gamma_{\rm ast}/dt_{\rm ast}$ against $t_{\rm ast}$ for the cases considered in previous sections (QSO-disk, QSO-bulge, disk-disk, bulge-disk, and bulge-bulge lensing).
All of the results presented in figure 7 are calculated for ($l$, $b$)=(0$^\circ$, 0$^\circ$).
For QSO-disk and QSO-bulge cases, there is a sharp peak at short duration ($\sim$ a few years), which corresponds to events caused by the stars located at the lower end of the stellar mass function.
Thus, if one can trace this peaks, it is possible to determine {`}typical{'} lens mass, which is likely to be (or at least fairly close to) the minimum cutoff mass of the mass function (provided a sharp cutoff of the stellar mass function really exists).
For instance, for QSO-disk lensing cases the tangential velocity of lens is likely to be around 440 km/s (see figure 2) and the event duration becomes independent of the lens distance $D_{\rm d}$ (equation [\ref{eq:t-ast-av2}])
.
Thus the strong peak of the differential event rate is likely to be caused by lenses with rather unique parameters; $m\sim m_{\rm L}$ and $v_\perp = 440$ km/s.
For bulge-disk and bulge-bulge cases, there still exist a peak like QSO-disk/bilge lensing cases, but at $t_{\rm ast}\sim 0$.
Also, the probability of lensing is lower by more than an order of magnitude.
Hence, it is not easy to estimate simply a 'typical' lens mass for bulge source cases but one has to take a careful statistical treatment.
For disk-disk cases, there is no such a sharp peak in $d\Gamma_{\rm ast}/dt_{\rm ast}$ plot.
Possible reason for this is that for disk-disk case the lens distance widely ranges ($(D_{\rm d}/D_{\rm s})$ from 0 to unity) and also the tangential velocity varies significantly with lens distance (see figure 2), which causes nearly equal-weight contribution of lens with wide range of $v_\perp$ and $D_{\rm d}$.

Figure 8, which is a plot of cumulative event rate vs event duration, shows this situation from a different point of view.
In figure 8, for QSO-disk and bulge-bulge events, the cumulative event rate is steeply rising with increasing even duration, demonstrating that more than 80\% of events have event duration less than 4 yr.
For bulge-disk and QSO-bulge lensing cases, the rise at short event duration is relatively gentle, but yet 80\% of events have duration less than 8 yr.
In contrast to these cases, the cumulative event rate for disk-disk lensing case is only slowly increasing with event duration, indicating considerable contributions of events with long duration.

Note that for QSO-disk and bulge-bulge lensing cases the steep rise in the cumulative event rate strongly depends on whether a sharp truncation of the stellar mass function really exists.
If the stellar mass function has a turnover at a certain mass and gently declining toward its lower end,  then the rise of the cumulative event rate would be softened.
In that case, it is not so easy to estimate of a typical mass of {`}dominant{'} population.
Alternatively, one needs to calculate many of plots like figure 8 with various model parameters as templates, and a comparison between these templates and observed event durations should be made to extract reliable information on the lens mass (and possibly the stella mass function) from astrometric microlensing events.
Thus, figure 8 gives an example of such templates to be compared with observed astrometric events.

\section{Discussion and Conclusion}

In this section, we discuss the implication of astrometric microlensing for VERA by focusing on the strategy of practical observations.
First, here we briefly summarize advantages as well as disadvantages of using VERA for astrometric microlensing search.
Advantages of the distant source case (which is the case for VERA) can be summarized as follows: I) large event probability expected from the optical depth and event rate, II) short event duration compared to the mission lifetime, and III) less uncertainty in lens mass determination, as the expected event duration is independent of the lens distance.
On the other hand, disadvantages of distant source observation with VERA are: I) small number of sources compared to stars in the Galaxy, and II) possible structural variation of the sources which could cause an apparent position shift of sources.

As for the first disadvantage (shortage of sources), it is true that the number of radio sources is fairly small compared to the number of visible stars in the Galaxy.
For instance, the number of compact radio sources that are expected to be observable with VERA is around 2000 (e.g., Ma et al. 1998, Peck \& Beasley 1998).
Moreover, the number density of compact sources in the Galactic plane is smaller than that in the off-plane region, since surveys of compact sources are conducted mainly in the off-plane regions (e.g., Patnaik et al. 1992; Peck \& Beasley 1998).
Thus, in order to have a sufficient number of sources, one has to conduct a survey in the Galactic plane region (e.g., Honma et al. 2000).

However, even at this stage there exist some sources which can be used for astrometric microlensing search.
For instance, there is a radio source only 0.7$^\circ$ away from the Galactic center (Backer \& Sramek 1999; Reid et al. 1999).
Also, according to Lazio \& Cordes (1998) there are at least five extra-galactic radio sources within a few degrees from the Galactic center (including the one 0.7$^\circ$ away from the Galactic center).
Properties of the five sources are listed in table 3, including event rates for disk lens and bulge lens cases.
The event rates for disk lens cases vary from $6.0\times 10^{-3}$ to $1.1\times 10^{-2}$ event/yr, and the total event rate for the five sources is found to be $4.0\times 10^{-2}$ event/yr.
Hence, even if one monitors only these five sources, one would detect an astrometric microlensing event caused by a disk star within 25 yr.
The event rate for bulge lens is also quite high, varying from 2.8 to 4.7 $\times 10^{-3}$ event per year.
If the bulge event rate is included, the total event rate for the five sources is $6.0\times 10^{-2}$ event/yr, indicating that within 16 years at least one of the sources is being lensed by bulge or disk star.
Note that these five sources are located in about $2^\circ \times 8^\circ$ area around the Galactic center (see figure 2 in Lazio \& Cordes 1998).
If the same number density applies to the rest of the Galactic plane, one can expect to find fairly large number of radio sources (i.e., a few tens).
Therefore, the shortage of sources can be resolved by further search for radio sources in the Galactic plane.

As to the second disadvantage (QSO structure variation), it is not easy to distinguish the structure effect and astrometric microlensing within a short period.
However, the image trajectory of an astrometric microlensing is a perfect circle for most cases except for events with extremely small impact parameter.
On the other hand, it is unlikely that an image motion caused by the structure variation becomes a perfect circle.
Hence, if one monitors the source position for the full event duration, one can easily discriminate the structural effect and astrometric microlensing (but of course there is no way to separate both effects if the two effects simultaneously happen to one source, and to avoid this it is better not to observe sources with significant structure variation).

In conclusion, astrometric microlensing events due to stars are practically detectable based on the observation of distant radio sources with VERA.
If such events are detected, one can estimate the lens mass, and this will probably brings us new information on the stellar mass function at the lower end.
According to the event rate calculation presented in the previous sections, it is not easy to detect an event, but it is possible if one monitors a few tens of sources for a decade or so.
Although this is time-consuming, it is a worthwhile study when one considers its scientific importance.

\newpage
\section*{References}
\def\re{\hangindent=1pc \noindent}

\re Backer, D., \& Sramek, R.A. 1999, ApJ, 524, 805

\re Boden, A. F., Shao, M., \& van Buren, D. 1998, ApJ, 502, 538

\re Dehnen W., \& Binney J. 1998, MNRAS, 294, 429

\re Dominik, M., \& Sahu, K.C. 2000, ApJ, 534, 213

\re Douglas J.N., Bash F.N., Bozyan F.A., Torrence G.W., Wolfe C. 1996, AJ 111, 1945 (Texas survey)

\re Griest, K. 1991, ApJ, 366, 412

\re H\o{}g, E., Novikov, I. D., \& Polnarev, A. G. 1995, A\&A, 294, 287

\re Honma, M., Kawaguchi, N., \& Sasao, T. 2000, in Radio Telescope, ed. H. R. Buthcer, Proc. SPIE 4015, 624

\re Honma, M. et al. 2000, PASJ, 52, 631

\re Honma, M. 2001, PASJ, 53, 233

\re Hosokawa, M., Ohnishi, K., Fukushima, T., \& Takeuti, M. 1993, A\&A, 278, L27
\re Hosokawa, M., Ohnishi, K., \& Fukushima, T. 1997, AJ, 114, 1508

\re Kerr, F.J., \& Lynden-Bell, D. 1986, MNRAS, 221, 1023

\re Lazio T.J.W., Cordes, J.M. 1998, ApJS, 118, 201

\re Lynden-Bell D., 1962, MNRAS, 123, 447

\re Ma, C., Arias, E.F., Eubanks, T.M., Fey, A.L., Gontier, A.M., Jacobs, C.S., Sovers, O.J., Archinal, B.A., \& Charlot, P. 1998, AJ, 116, 516 (ICRF catalog)

\re Miralda-Escude, J. 1996, ApJ, 470, L113

\re Miyamoto, M., \& Yoshi, Y. 1995, AJ, 110, 1427

\re Paczy\'{n}ski, B. 1986, ApJ, 304, 1

\re Paczy\'{n}ski, B. 1996, Acta Astronomoica, 46, 291

\re Paczy\'{n}ski, B. 1998, ApJ, 494, L23

\re Patnaik, A.R., Browne, I.W.A., Wilkinson, P.N., \& Wrobel, J.M. 1992, MNRAS, 254, 655

\re Peck, A.B., \& Beasley, A.J. 1998, in Radio Emission from Galactic and Extragalactic Compact Sources, IAU Colloquium 464, ed. J.A. Zensus, G.B. Taylor, J.M. Wrobel, ASP Conf. Ser. 144, 155 (VLBA calibrator survey)

\re Reid, M.J., Readhead, A.C.S., Vermeulen, R.C., Treuhaft, R.N. 1999, ApJ, 524, 816

\re Tiede G.P., Terndrup D.M., 1999, AJ, 118, 895

\re Sasao, T. 1996, in Proc. of 4th Asia-Pacific Telescope Workshop, ed. E. A. King, 94 (Sidney, Australia Telescope National Facility)

\re Walker, M. A. 1995, ApJ, 453, 37

\clearpage
\begin{center}
Table 1. Summary of the mass model of the Galaxy used in this paper.
\vspace{1cm}

\begin{tabular}{ccc}
\hline\hline
component & parameter & value \\
\hline
bulge & total mass ($M_{\rm b}$) & 8$\times 10^{9} M_\odot$ \\
      & scale length ($a$) & 1 kpc \\
disk  & local density ($\rho_{\rm d}$) & 0.08$M_\odot$/pc$^3$ \\
      & scale length ($d$) & 3.5 kpc \\
      & scale height ($h$) & 300 pc \\
      & velocity dispersion ($\sigma_{\rm d}$) & 20 km/s \\
\hline
\end{tabular}
\end{center}

\clearpage
\begin{center}
Table 2. Summary of source and lens pair considered in this paper.
\vspace{1cm}

\begin{tabular}{cccc}
\hline\hline
name in text & source & lens & $D_{\rm s}$ \\
\hline
QSO-disk & QSO & disk star & $\infty$ \\
QSO-bulge& QSO & bulge star& $\infty$ \\
disk-disk& disk star & disk star & 8 kpc \\
bulge-disk& bulge star & disk star & 10 kpc \\
bulge-bulge& bulge star & bulge star & 10 kpc \\
\hline
\end{tabular}
\end{center}

\clearpage
\begin{center}
Table 3. Possible targets for astrometric microlensing search with VERA.
\vspace{1cm}

\begin{tabular}{ccccc}
\hline\hline
source name$^{\rm a}$ & $l$ & $b$ & $\Gamma_{\rm disk}$ (event/yr) & $\Gamma_{\rm bulge}$ (event/yr) \\
\hline
TXS 1741-312 & 357.865 & $-$0.996 & $6.0\times 10^{-3}$ & $3.9\times 10^{-3}$\\
TXS 1740-309 & 358.002 & $-$0.637 & $7.5\times 10^{-3}$ & $4.1\times 10^{-3}$\\
TXS 1739-297 & 358.917 & +0.072 & $1.1\times 10^{-2}$ & $4.7\times 10^{-3}$\\
	& 358.983 & +0.580 & $7.9\times 10^{-3}$ & $4.7\times 10^{-3}$\\
TXS 1748-253 & 3.745 & +0.635 & $7.3\times 10^{-3}$ & $2.8\times 10^{-3}$\\
\hline
total ($\sum \Gamma$) & & & $4.0\times 10^{-2}$ & $2.0\times 10^{-2}$\\
\hline
\end{tabular}
\end{center}
\vspace{1cm}

$^{\rm a}$ Source name in the Texas survey (Douglas et al. 1996).
In Lazio \& Cordes (1998) source names are given by the galactic longitude and latitude.

\clearpage
\begin{figure}[t]
\vspace{16cm}
\epsfxsize=150pt
\epsfbox[0 50 200 150]{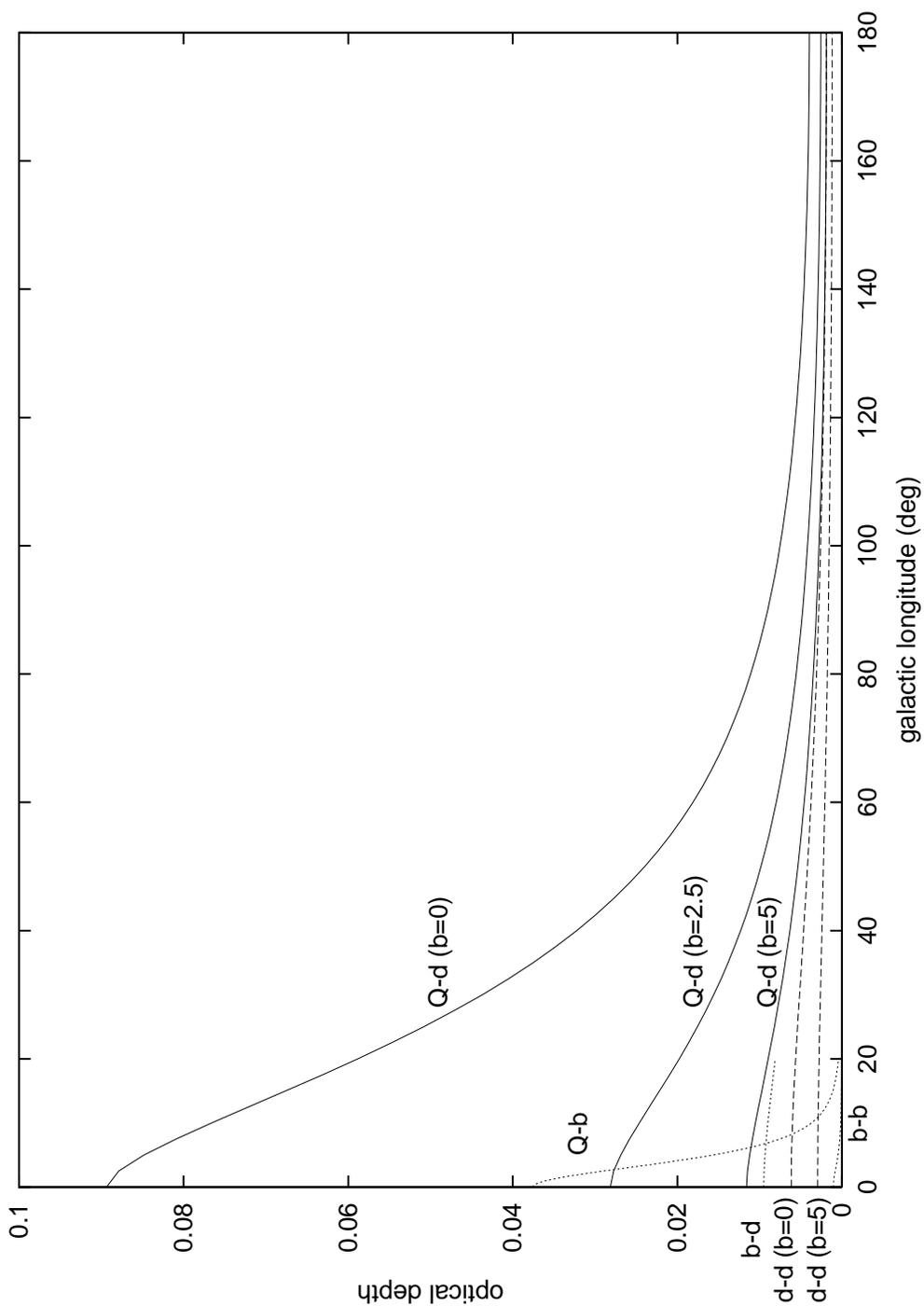}
\caption{Optical depth for QSO-disk, QSO-bulge, disk-disk, bulge-disk, and bulge-bulge lensing cases.
Thin lines are optical depth for QSO-disk lensing for $b=0^\circ$, $b=2.5^\circ$, and $b=5^\circ$, respectively.
Dashed lines are for disk-disk lensing (for $b=0^\circ$, and $b=5^\circ$).
Dotted lines are for QSO-bulge lensing, for bulge-disk lensing, and for bulge-bulge lensing cases from top to bottom.
}
\end{figure}

\begin{figure}[t]
\vspace{16cm}
\epsfxsize=150pt
\epsfbox[0 50 200 150]{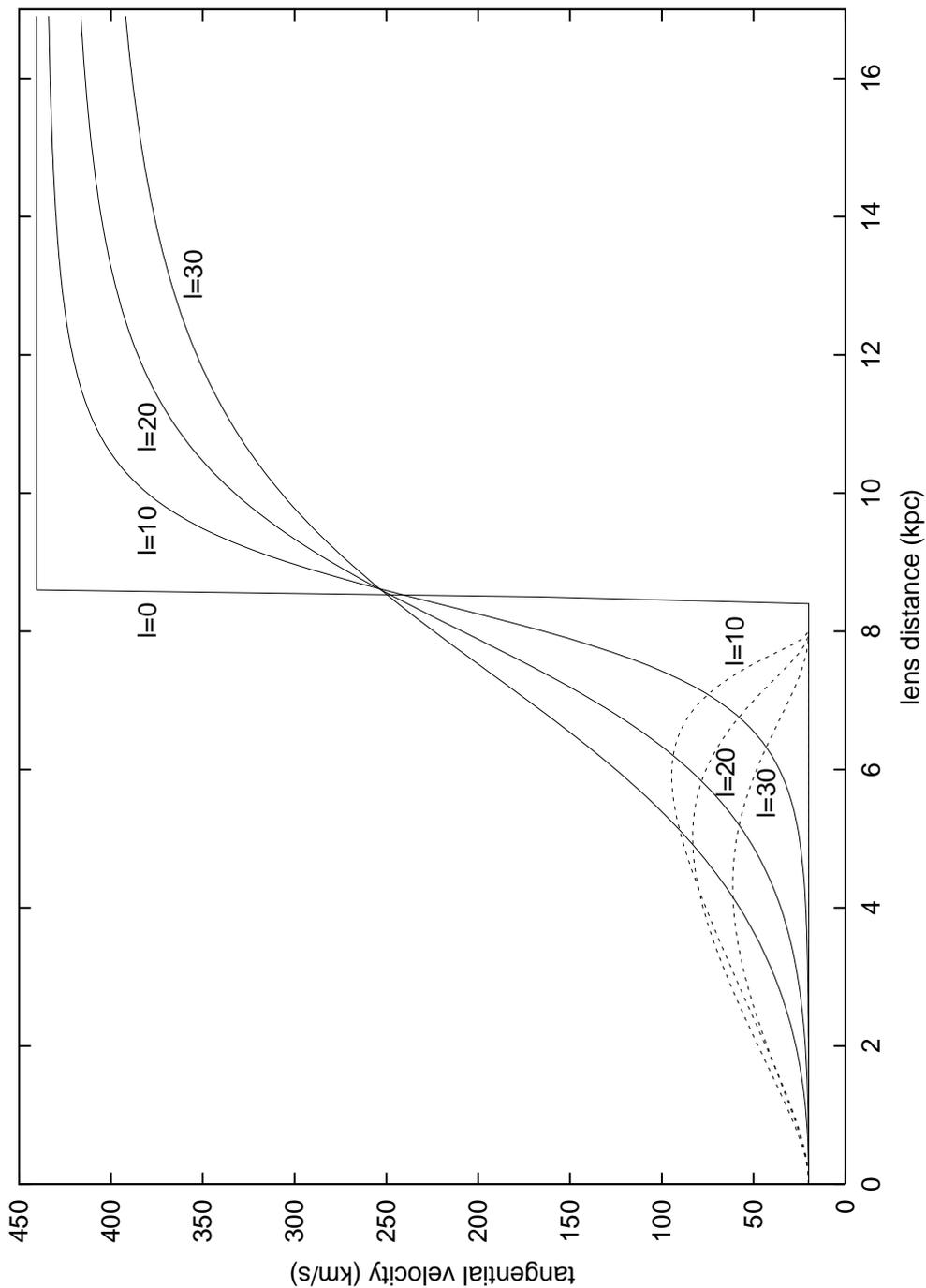}
\caption{Distribution of tangential velocity of the lens relative to microlensing tube for $l=0^\circ$, $10^\circ$, $20^\circ$, and $30^\circ$ cases (thin lines for QSO-disk lensing cases, and dashed lines for disk-disk lensing cases).
For QSO-disk lensing, the tangential velocity becomes close to 2$v_{\rm circ}$ ($\sim 440$ km/s) behind the Galactic center.
}
\end{figure}

\begin{figure}[t]
\vspace{16cm}
\epsfxsize=150pt
\epsfbox[0 50 200 150]{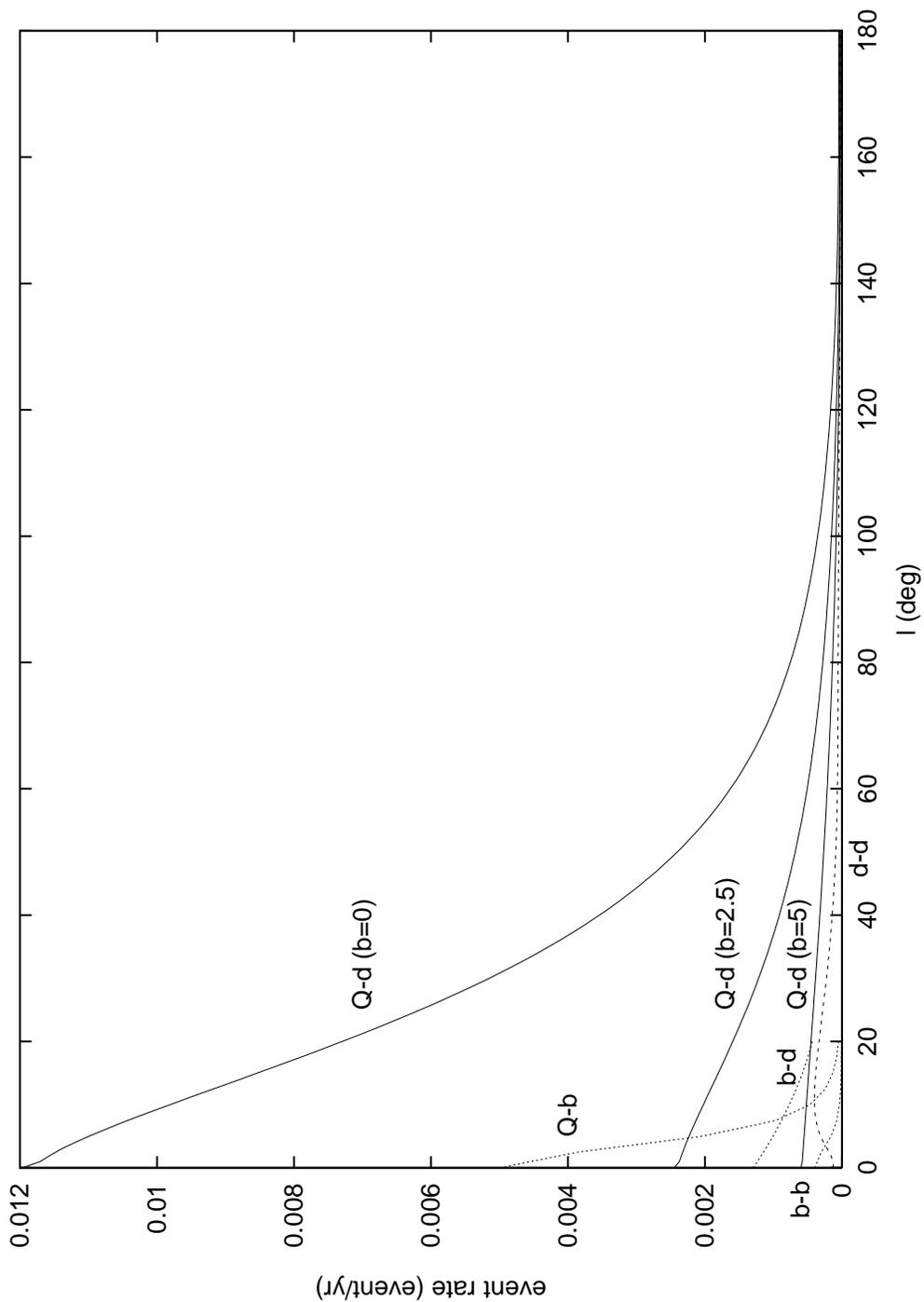}
\caption{Distribution of event rate with galactic longitude $l$ (notations are the same to figure 1).
For QSO-disk lensing (thin lines), three cases for $b$ ($0^\circ$, $2.5^\circ$, and $5^\circ$) are shown.
While the event rate for QSO-disk/bulge lensing case becomes maximum toward the Galactic center, the event rate for disk-disk lensing (dashed lines) is largest around $l=10^\circ$.
}
\end{figure}

\begin{figure}[t]
\vspace{16cm}
\epsfxsize=150pt
\epsfbox[0 50 200 150]{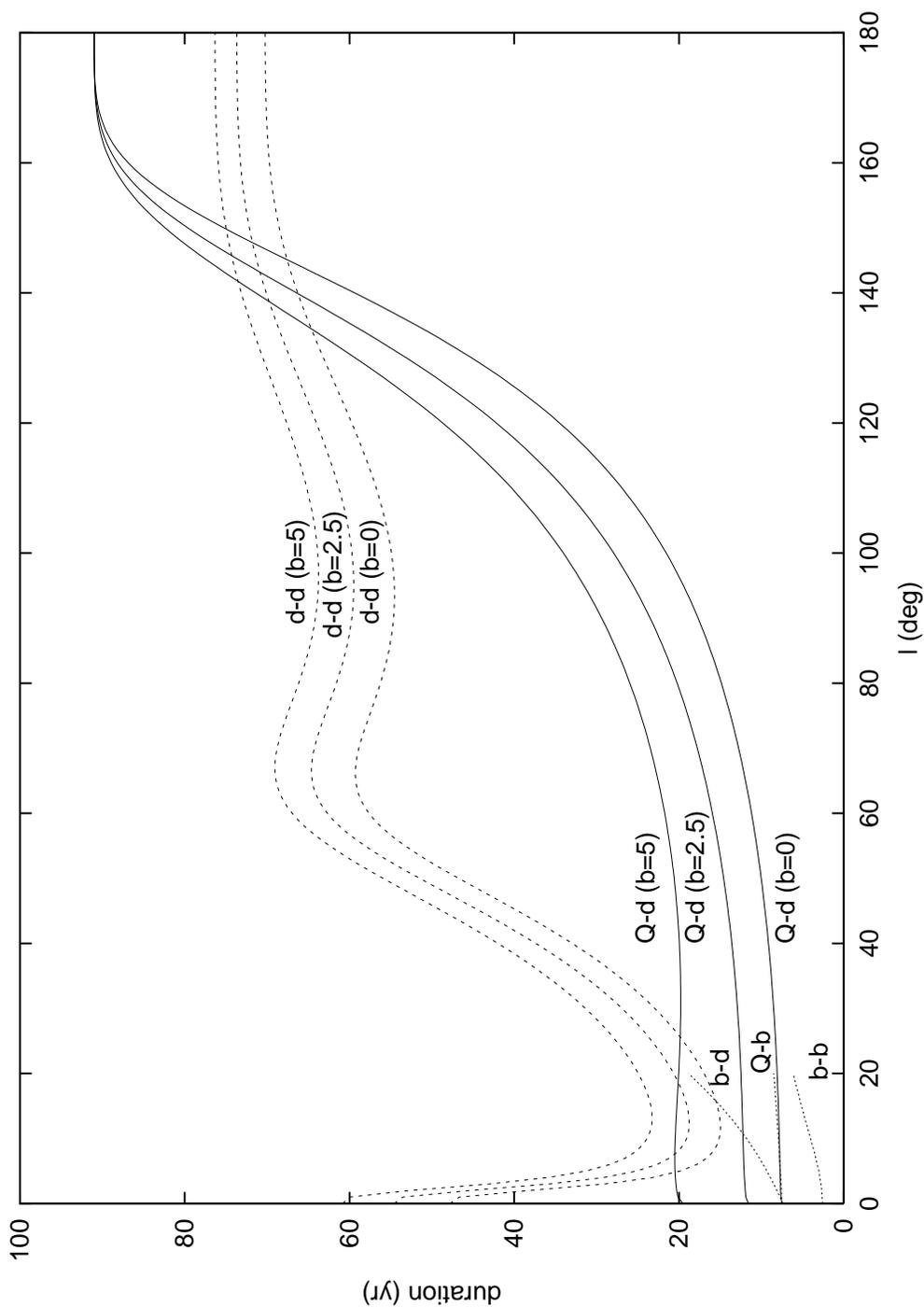}
\caption{Distribution of expected event duration ($\langle t_{\rm ast} \rangle$) for $b=0^\circ$, $b=2.5^\circ$, and $b=5^\circ$, for both QSO-disk (thin lines) and disk-disk lensing cases (dashed lines).
Also shown are event duration distributions for QSO-bulge lensing, bulge-disk lensing, and bulge-bulge lensing (dotted lines, shown only for $l$ within $20^\circ$).
}
\end{figure}

\begin{figure}[t]
\vspace{16cm}
\epsfxsize=150pt
\epsfbox[0 50 200 150]{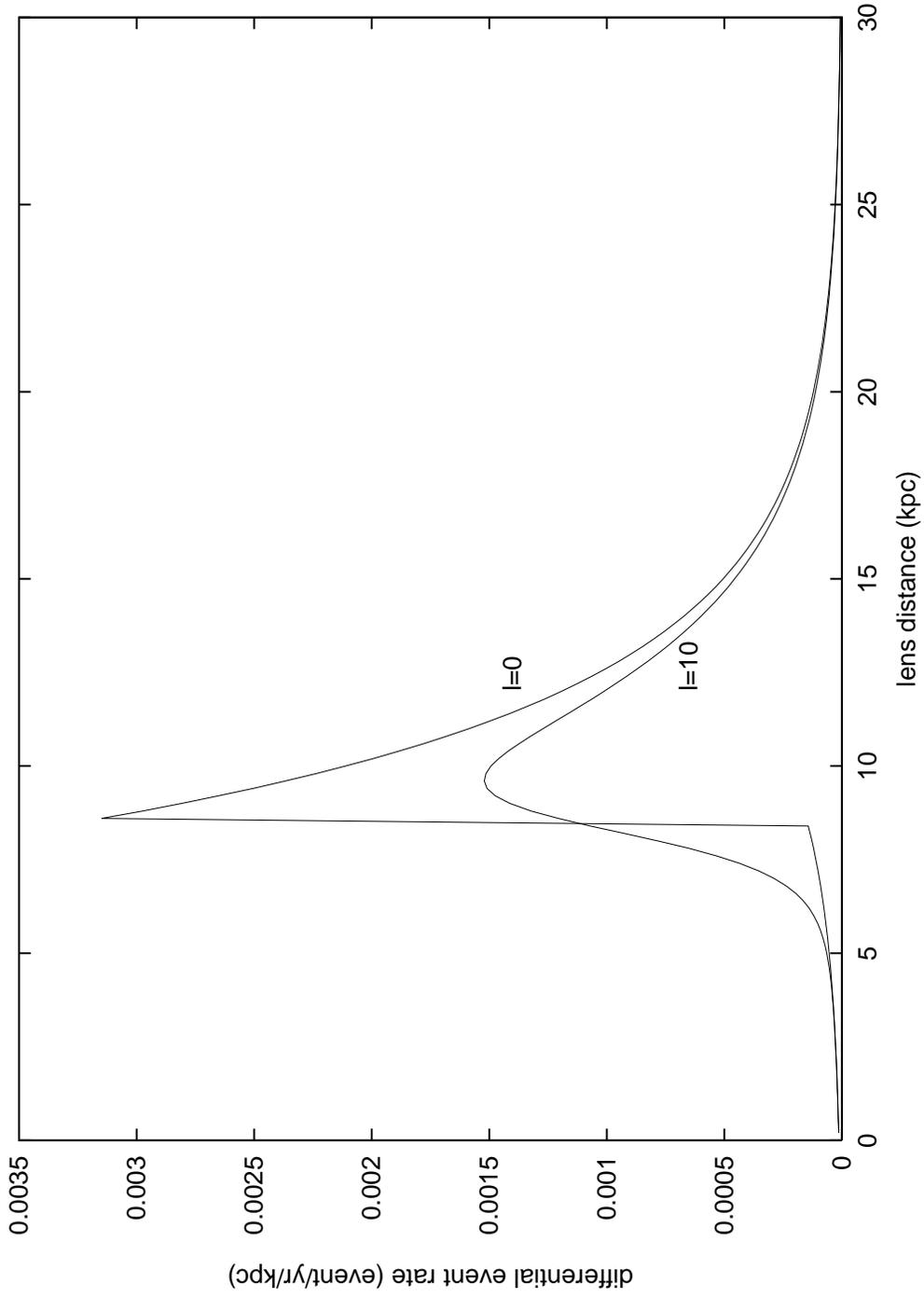}
\caption{Distribution of differential event rate with lens distance for QSO-disk lensing case (for $l=0^\circ$ and $10^\circ$).
The differential event rate sharply peaked behind the galactic center.
}
\end{figure}

\begin{figure}[t]
\vspace{16cm}
\epsfxsize=150pt
\epsfbox[0 50 200 150]{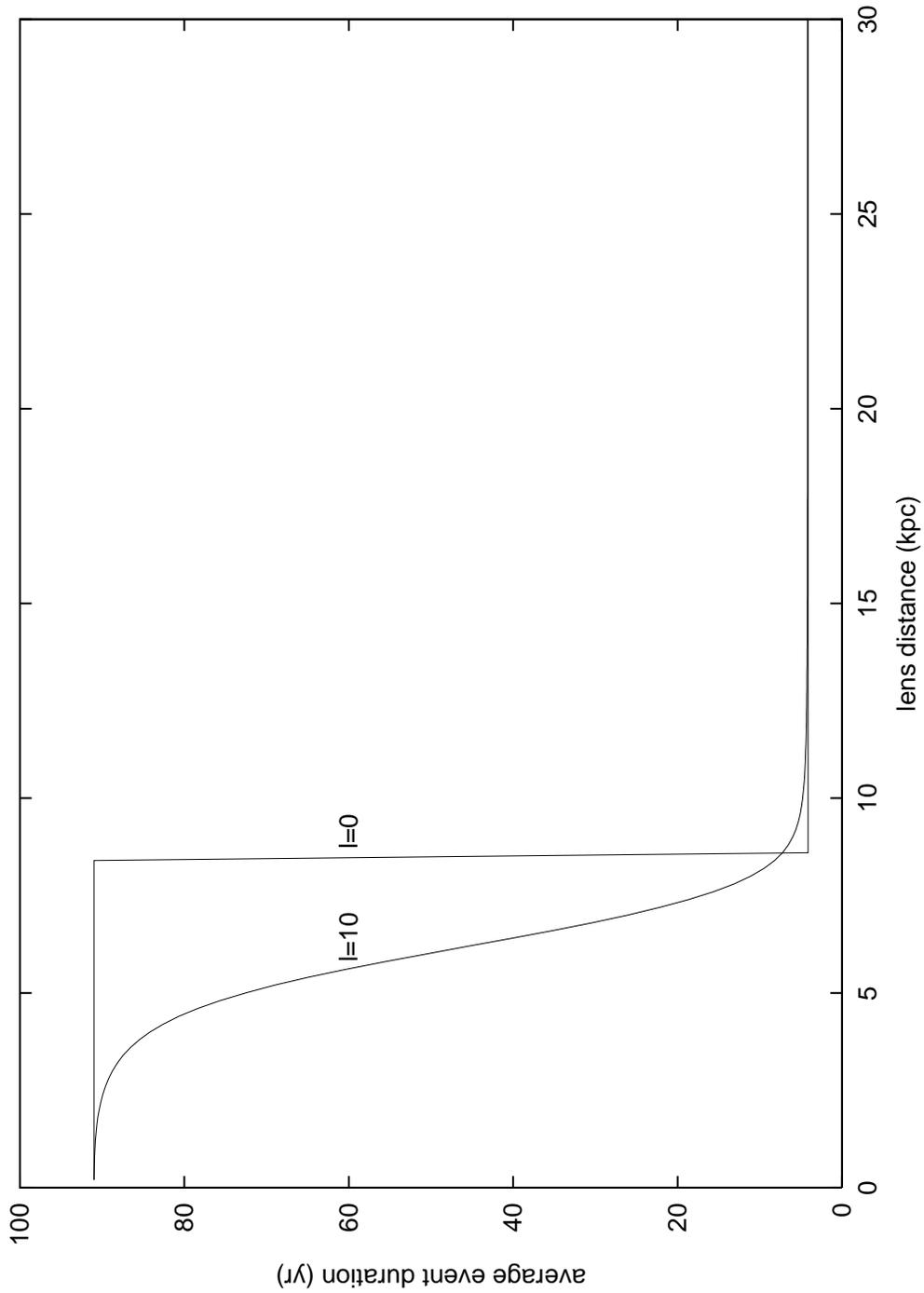}
\caption{Distribution of averaged event duration with lens distance for QSO-disk lensing case (for $l=0^\circ$ and $10^\circ$).
The disk stars behind the galactic center cause events with short duration ($\sim $ 4 yr), while those near the Sun cause much longer event ($\sim$ 90 yr).
}
\end{figure}

\begin{figure}[t]
\vspace{16cm}
\epsfxsize=150pt
\epsfbox[0 50 200 150]{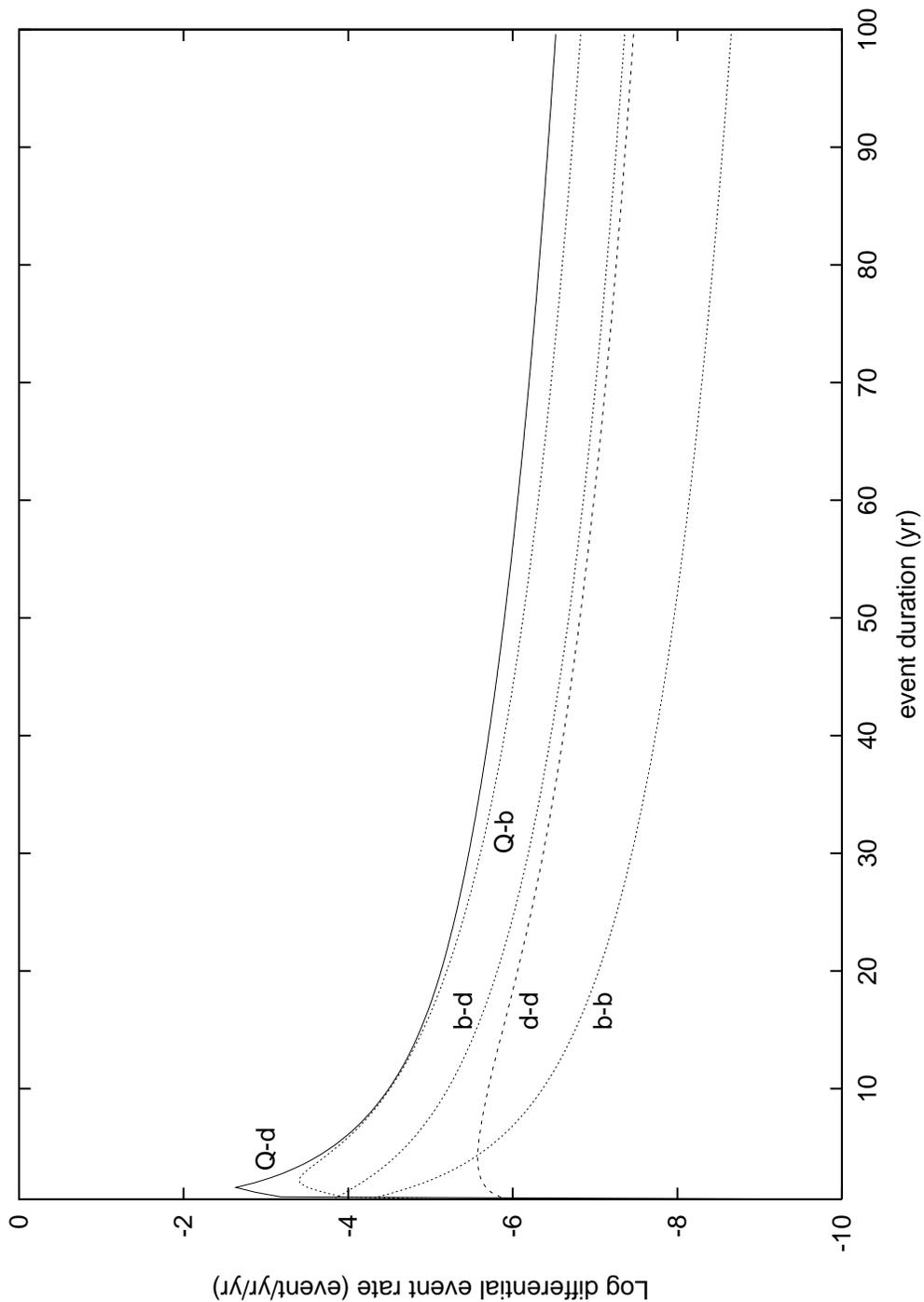}
\caption{Distribution of differential event rate ($d\Gamma_{\rm ast}/t_{\rm ast}$) against event duration $t_{\rm ast}$ (for $l=0^\circ$ and $b=0^\circ$).
For QSO-disk/bulge lensing cases, the differential event rate has a sharp peak at short duration of a few years, while that for disk-disk lensing is rather flat.
}
\end{figure}

\begin{figure}[t]
\vspace{16cm}
\epsfxsize=150pt
\epsfbox[0 50 200 150]{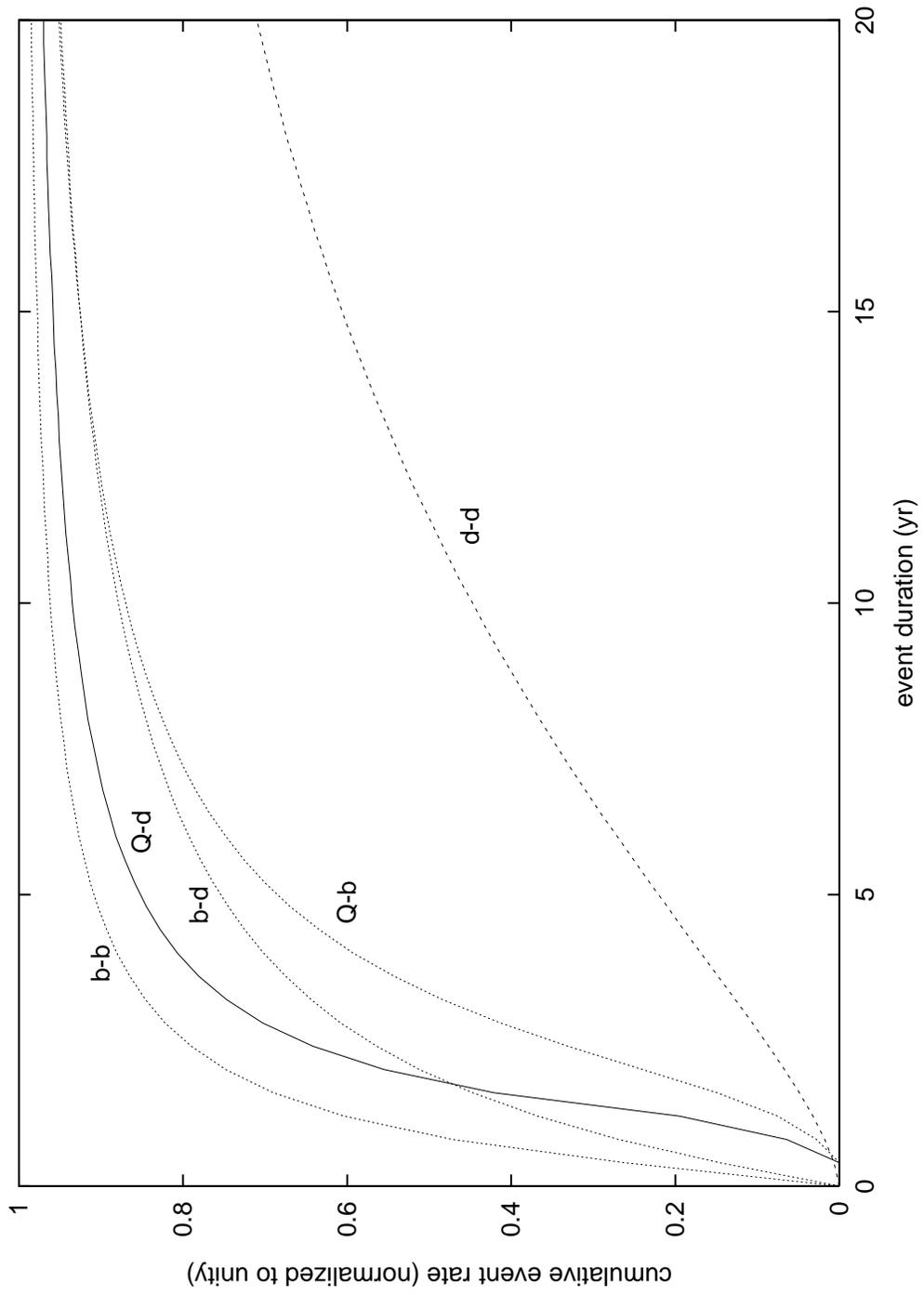}
\caption{Cumulative event rate against event duration for QSO-disk, QSO-bulge, disk-disk, bulge-bulge and bulge-disk lensing cases.
QSO-disk and bulge-bulge cases show steep rises of cumulative event rate at short event duration, in contrast to disk-disk lensing case.
}
\end{figure}

\end{document}